# Neural population geometry: An approach for understanding biological and artificial neural networks

**SueYeon Chung[1][†] and L. F. Abbott[1]**

[1] Center for Theoretical Neuroscience, Columbia University, New York City, United States
† Corresponding author: sueyeon.chung@columbia.edu

**Highlights**

- Manifold-like representations arise when a set of neurons in a biological or artificial neural network exhibits variability in response to stimuli or through internal recurrent dynamics.
- Approaches focused on analyzing geometric properties of neural populations, i.e. *neural population geometry*, have emerged as a promising population-level analysis technique connecting neural responses and task implementation.
- We highlight recent studies of neural population geometry: untangling in perception, classification theory of manifolds, abstraction in cognitive systems, topology underlying cognitive maps, dynamic untangling in motor systems, and a dynamic approach to cognition.
- Future directions include developing geometric measures as a population-level hypothesis, connecting representational geometry to biophysical properties of neurons, developing theories of neural population geometry for a larger array of tasks.

**Abstract**

Advances in experimental neuroscience have transformed our ability to explore the structure and function of neural circuits. At the same time, advances in machine learning have unleashed the remarkable computational power of artificial neural networks (ANNs). While these two fields have different tools and applications, they present a similar challenge: namely, understanding how information is embedded and processed through high-dimensional representations to solve complex tasks. One approach to addressing this challenge is to utilize mathematical and computational tools to analyze the geometry of these high-dimensional representations, i.e., neural population geometry. We review examples of geometrical approaches providing insight into the function of biological and artificial neural networks: representation untangling in perception, a geometric theory of classification capacity, disentanglement and abstraction in cognitive systems, topological representations underlying cognitive maps, dynamic untangling in motor systems, and a dynamical approach to cognition. Together, these findings illustrate an exciting trend at the intersection of machine learning, neuroscience, and geometry, in which neural population geometry provides a useful population-level mechanistic descriptor underlying task implementation. Importantly, geometric descriptions are applicable across sensory modalities, brain regions, network architectures, and timescales. Thus, neural population geometry has the potential to unify our understanding of structure and function in biological and artificial neural networks, bridging the gap between single neurons, population activities, and behavior.

## Introduction

Neural circuits and artificial neural networks (ANNs) process information by constructing and manipulating highly distributed representations [1–4]. Patterns of activity in these systems, across either neurons or units, correspond to manifold-like representations (Box 1) - lines [5], surfaces [6,7], trajectories [8–10], subspaces [11], and clouds of points [12,13] - in a high dimensional 'neural state space,' where coordinates represent the activities of individual neurons or units. Approaches focused on studying geometric properties of these manifolds are becoming more widely used as advances in experimental neuroscience expand our ability to probe large neural populations [14], and advances in ANNs [15,16] introduce new challenges of interpretation.

In neuroscience, driven by advances in recording techniques, mainstream analysis tools have subsequently transitioned from single-neuron approaches [17,18] to population-level frameworks [1–3,19,20] that quantify and decode information represented across many neurons. Challenges arise when we consider large neural populations involved in complex tasks, as neurons often show mixed selectivity, i.e., selectivity to multiple coding variables [21], and real-world tasks often require robustness to nontrivial variability [6], precluding simplistic tuning-based analyses. The geometric analysis provides an approach suitable for addressing these challenges.

Since a number of large-scale task-optimized ANNs have outperformed traditional neuronal models in accounting for neural activity [22,23], ANNs have become a promising model system for studying neural circuits. One often-heard objection to the use of ANNs in modeling neural circuits is that ANNs merely replace one complicated system with an equally complicated system[24]. Indeed, the challenges in interpreting high-dimensional ANNs, containing millions of parameters, and neural populations are shared [25]. This highlights the need for powerful population-level tools that reveal mechanisms underlying neural network function. From this perspective, ANNs can serve as a testbed for developing population-level analysis techniques, such as geometric approaches, even if they are ultimately aimed at neuroscience applications.

In this review, we highlight important examples of how geometrical techniques and the insights they provide have aided the understanding of biological and artificial neural networks. We begin with an overview of recent theoretical developments linking neural population geometry to categorization capacity. We then discuss theoretical work on characterizing representational geometries across tasks and modalities, such as recognition and prediction in the sensory domain (perceptual untangling) and abstraction in the cognitive domain (disentanglement). We also discuss sensory or behavioral state transitions in the head direction system and hippocampus (topology discovery). Finally, we provide examples for which dynamical analysis of neural population geometry sheds light on representations in motor control (dynamic untangling) and complex cognitive tasks such as Bayesian inference.

> "***Manifold***" means a topological space that locally resembles a Euclidean space in mathematics. The term "neural manifold" has been used to refer to a broad set of geometric structures in neural population activity underlying various cognitive tasks, even though these population structures in real neural data are often no longer technically "manifolds" in a mathematical sense, mainly due to the presence of neural noise and but also often due to the sparse input sampling.
>
> ***Object manifolds*** [6,7,13] or ***perceptual manifolds*** [5,12] refer to sensory neurons' population structures that arise as a result of identity-preserving variabilities in the input stimulus space. The term ***neural manifold*** has been used more broadly to refer to low-dimensional subspaces underlying population activities embedded in high-dimensional neural state space, not only in (aforementioned) sensory brain regions but also in motor and cognitive brain regions [11,26,27].
>
> ***Point-cloud manifolds***: a point-cloud with an underlying manifold structure, where the typical source of the underlying manifold variability is stimulus variability (e.g., orientation or position) or neuronal variability (e.g., the shape of a neuron's tuning curve). Despite implied underlying manifold structure, the data often manifest themselves as point clouds, due to the sparse sampling of data from the available range of the stimulus/neuronal variabilities, and/or due to noise (input noise or stochastic neuronal noise).
>
> ***Neural population geometry*** refers to the configurations of these ***neural manifolds***, embedded in ambient ***neural state space***.
>
> Box 1. Clarifications on the use of the term "manifold" in neuroscience

## The Geometry of Perception and Decision Making

### Perceptual Untangling

It has been hypothesized that the role of ventral visual stream processing is to transform the representations of visual objects so that they become 'untangled', meaning that they are transformed into a form that is linearly separable [6,7] (Fig 1a). The concept of linear separability goes back to the early days of ANNs [28,29], and it still plays a central role in the analysis of neural population geometry. A task in which a subject must divide a large set of stimuli into two categories requires the *separation* of the neural activity patterns evoked by these stimuli into two sets corresponding to the two categories. We know from machine learning that this discriminability can be achieved easily if a hyperplane can separate the two sets of activities. Such a representation is called linearly separable. If, instead, the separating surface must be curved, dividing the two sets of neural population activities is more difficult. This insight is central to a number of the approaches we discuss.

The idea of untangling has been extended into the time domain [7,30]. In this case, neural population activity corresponds to a trajectory through neural state space. These studies posit that, at a given point in time, it is easier to predict future neural activity if this trajectory is straight than if it is convoluted. This led to the hypothesis that visual processing also serves to straighten temporal response trajectories [7]. This 'temporal straightening hypothesis' has been tested by measuring the curvature of the neural trajectory of responses to natural videos in neural network models and human perceptual space [30] (Fig 1b). Straightening of response trajectories occurs when natural video sequences, but not artificial video sequences, are presented.

## The Geometry of Abstraction

The principle of linear separability can also provide insight into more complex tasks beyond categorization.  Consider a task in which two sets of stimulus-response pairings, set A and set B, must be learned.  The task involves uncued 'context' switches between the use of set A and set B. An efficient solution is to represent the stimulus-response pairings in such a way that a transition between contexts can be accomplished by the rotation and/or translation of a dividing surface in the neural state space (Fig 1c).  Recordings from the prefrontal cortex, hippocampus, and results from task-trained neural networks [31] all indicate the use of 'disentangled' representation, quantified by a geometric measure called the parallelism score.  These studies provide direct neural evidence on how two different contexts are involved in such a task, and thus probe the level of abstraction and type of strategy being used by the animals and machines.  An important idea here, which will reappear in another context in the next section, is that while abstraction is achieved, the representation does not simply discard information about other variables [31].

## Extensions from Points to Manifolds

In the research covered thus far, neural population activity during a task has been considered to be a point (in the case of static stimuli) or a one-dimensional trajectory (in the case of time-dependent stimuli) in the neural state space. However, the same stimulus shown repeatedly will not result in the same point in state space being occupied; instead, neuronal variability will cause the points from different trials to jitter. The result is that each stimulus corresponds not to a point but to a point cloud whose size and shape depend on the amplitude and form of the neuronal variability. Furthermore, the presence of other sources of variability introduces the need to cluster responses into point-cloud manifolds (Box 1).  For example, if we want to distinguish dogs from cats, we may want to group the responses to images for different viewing angles, sizes, and animal breeds into one dog manifold and one cat manifold (Fig 1d). In this perspective, the problem of invariant object discrimination becomes that of separating neural manifolds[12].

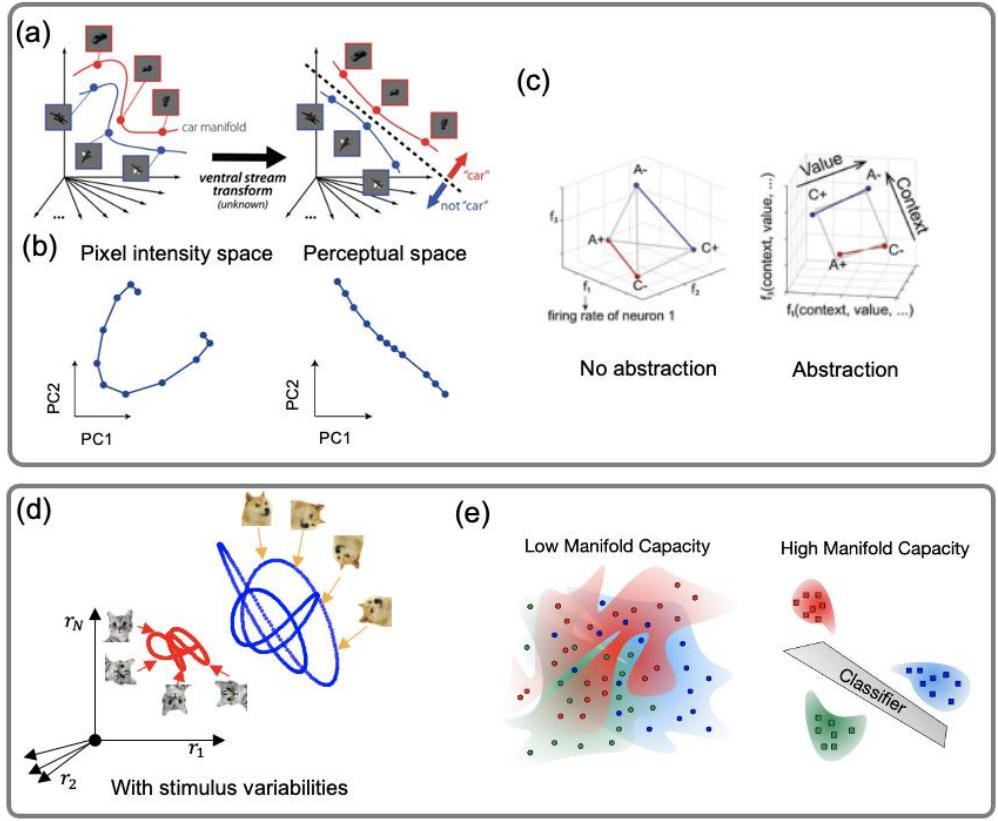

Fig 1. (a) Representation straightening for invariant object recognition (b) Temporal straightening for temporal natural video sequences. (c) Geometry of Abstraction. Representations encoding abstraction (i.e., cross-conditional generalization) show geometry where coding directions can be rotated or translated between conditions, known as parallelism (Right). (d) Neural manifolds arise as a result of stimulus variability. Population responses to two object classes (dog vs. cat) in the presence of the stimulus variability (orientation) gives rise to two object manifolds. Invariant object recognition becomes the problem of classifying between two object manifolds. Axes represent the firing rates of neurons. (e) Manifold capacity is high if object manifolds are well separated and low when object manifolds are entangled in neural state space. Part (a) adapted from [7]. Part (b) adapted from [30]. Part (c) adapted from [31]. Part (e) adapted from [32].

Determining the mechanism behind invariant object discrimination requires us to decipher how the structure across different instances of the same object are processed by the layers of the sensory hierarchy. This raises the question of how the structure of neural object manifolds is related to the separability of object categories. Theoretical work based on concepts from statistical physics has shown that linear separability of object manifolds, as defined by the object manifold capacity [12], a generalization of perceptron capacity, can be formally connected to the geometric properties of object manifolds such as their dimension, radius and correlation structure [12,13,33].

One result of this work is that the same level of linear separability can be achieved across different combinations of geometrical properties. For example, combinations of large/small dimensionality and small/large size of object manifolds can lead to similar capacities, because there is a tradeoff between the dimensionality and the radius of these manifolds. The untangling hypothesis can be extended to the idea that visual processing aims to provide well-separated manifolds that provide information about object identity while maintaining other image-related variables such as pose, position and scale (Fig 1a,d).

In this framework, the notion of manifold capacity has several interpretations. While the manifold capacity measures the linear separability of object classes, it also measures the storage capacity of object classes in a given representation (i.e., the maximum number of object classes that can be read out linearly). Small manifold dimensions and radii predict high manifold capacity and vice versa (Fig 1e). This theory has been used to show how categorical information emerges across layer hierarchy as a result of geometrical changes in ANNs implementing visual object recognition [13], speech recognition [32], and language prediction tasks [34]. These ANN models are known to have a high neural predictivity with corresponding brain regions in the macaque visual cortex [22,35], human auditory cortex [36], and language processing regions [37]. In addition, promising preliminary results in mouse and macaque visual cortex [38,39] show that this theory can also be used directly to characterize neural data. These examples demonstrate how the untangling hypothesis has motivated advancements in new theoretical frameworks, such as manifold capacity theory, allowing for a more refined geometric analysis of representations in biological and artificial neural networks.

## The Intrinsic Geometry of Representation

Another approach to understanding high-dimensional neural activity focuses on the observation that the neural activity lies on lower-dimensional subspaces, i.e., *neural manifolds* (Box 1). To understand the structure of these neural manifolds, many recent studies have employed various dimensionality reduction techniques to the analysis of neural data. Dimensional reduction refers to manipulations used to identify the shape, location, and orientation of neural data within the neural state space. Widely used linear methods such as principal components analysis (PCA) provide a Cartesian coordinate basis describing subspaces in which the data lie. It is also useful to determine the geometric properties that characterize the intrinsic space defined by the data, which, in general, requires nonlinear dimensionality reduction methods. To be concrete, consider the responses of a population of neurons to a set of stimuli described by two variables (disregarding neural noise for simplicity). We might assume that these data can be described as a function of these two stimulus variables. If this is indeed the case, the responses lie on a two-dimensional surface, but that surface is not necessarily a flat plane. In fact, the surface might be convoluted and lie in a considerably higher dimension. PCA will find this higher dimensional embedding space, whereas nonlinear methods can find the curved surface itself.

A large number of nonlinear dimensionality reduction methods are available, including Isomap [40], LLE [41], tSNE [42], MDS [43], PHATE[44] and UMAP [45]. Although powerful, these nonlinear methods assume that underlying manifolds are topologically simple and can fail to capture the neural manifold structure if the underlying topology is complex. Computational advances have been made in an effort to understand how brain regions encode directional or spatial information, such as the head direction system and the hippocampus. Chaudhuri et al. [46] utilized a technique known as Spline Parameterization for Unsupervised Decoding (SPUD) (Fig. 2a) to discover the ring structure underlying the mammalian head direction system. This technique uses an approach called persistent homology[47,48] in which persistent features determine the intrinsic dimension used to discover underlying non-trivial topological structure in the data.

Meanwhile, recent work in the hippocampus introduced a topologically motivated method called Manifold Inference from Neural Dynamics (MIND) [49,50] (Fig 2b) to characterize neural activity in the CA1 region of the hippocampus during a foraging and sound manipulation task. In MIND, distances between nearby states are defined by transition probabilities, which gives rise to the notion of intrinsic dimensions relevant for topological maps underlying task implementation.

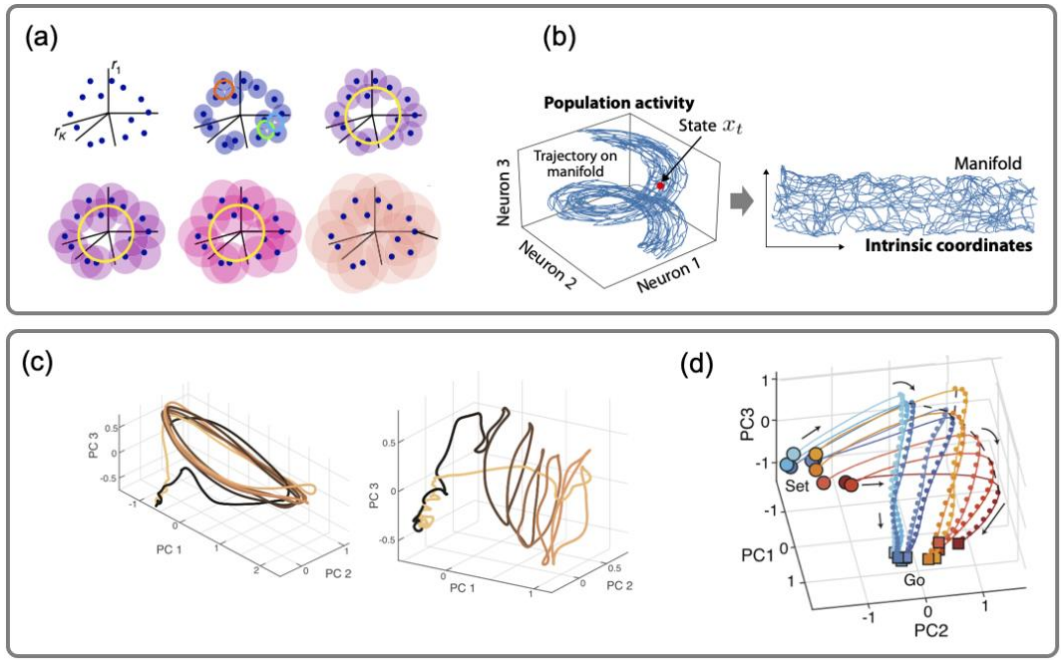

Fig 2. (a-b) Manifold discovery methods. (a) Spine Parameterization for Unsupervised Decoding (SPUD). (b) Manifold Inference for Neural Dynamics (MIND). (c-d) Population dynamics as cognition. (c) (Left) Temporal trajectories during macaque cycling task in M1 and (Right) SMA. (d) Dorsomedial Frontal Cortex (DMFC) response profiles during Bayesian computation. Part (a) adapted from [46]. Part (b) adapted from [10]. Part (c) adapted from [51]. Part (d) adapted from [8].

# The Geometry of Movement and Cognition

## Dynamic untangling of internally generated activities

The concept of untangling has also been applied to the neural trajectories recorded from the motor cortex during movement. In studies of motor regions, we are interested not only in how body movements are represented but, importantly, in how they are generated. How can we determine whether a given region of the brain is playing a significant role in movement generation as opposed to merely reflecting the effects of activity generated elsewhere? In a closed dynamical system, the rate of change of any dynamic variable is a function of all the other dynamic variables. Thus, it is impossible for a single point in the state-space to be associated with two different rates of change. This is equivalent to the statement that state-space trajectories in such a system cannot cross over themselves. A 'tangling index' has been introduced to identify cases when the trajectories of recorded populations of neurons actually or come close to crossing [9]. Using this measure, it was shown that tangling is much lower in the primary motor cortex during a cycling task than in areas such as the primary sensory cortex or in muscle activities during the same task. This supports the idea that the motor cortex acts as a generator, whereas activity in the somatosensory system and in muscles is a response to motor drive.

Interestingly, a follow-up study of neural activity trajectories during the cycling task in the supplementary motor area (SMA) provided geometric evidence of the well-known role of this area in motor sequencing [51]. Activity in the motor cortex repeated across cycles of the cycling task, but SMA activity followed a helical trajectory, providing s neural representation of the sequence of cycles made during the task (Fig 2c). A similar result was obtained in model recurrent neural networks; a helical representation arose when the network was required to keep track of the number of cycles it had generated [51]. These examples illustrate the extension of the use of geometric analyses, which we first discussed in relation to perception, to motor systems.

## Population dynamics as cognition

There is a long history of relating dynamic motifs in recurrent networks to cognitive functions: fixed points and memory [52,53], line attractors and integration [54], and limit cycles with various neuronal oscillation patterns [55]. These ideas have more recently been extended to a general program linking dynamics to cognition [56,57].

For example, work in the macaque frontal cortex during a time reproduction task, which requires subjects to reproduce the duration of a time interval, demonstrated that experience warps neural population representations[8]. This mechanism allows for the incorporation of prior statistics in the map from sensory representation to motor output [8]. A geometric analysis of the activity in recurrent neural networks trained to perform this task revealed how curvature supports an underlying Bayesian computation (Fig 2d).

# Conclusion

The neural population geometry approach suggests many open problems and future opportunities at the intersection between neuroscience and artificial intelligence. Below we enumerate opportunities and challenges for future study.

First, the neural population geometry can serve as a more accurate population-level descriptor compared to simple task-level probes, as representations with the same level of task capacity can have different geometric configurations [12,33]. Notably, dimensionality [58] is an important population-level metric capturing task information and representational redundancy. This can be further extended with other complementary measures necessary for a full understanding of computation. For example, invariant object classification capacity [12,13] is determined not only by an object manifold's dimension but also, crucially, by its radius.

Second, as the list of tasks and brain regions showing interesting population geometric structure is growing at a rapid pace [26,59–63], future theoretical developments may need to address the formal connection between representational geometric properties and the encoded task information for a larger array of tasks.

Third, future directions should include uncovering the relationship between population geometry and specific biophysical properties of neurons. In the neural geometry underlying Bayesian computation [8], the curvatures of trajectories are linked to the distributions of priors encoded by each neuron. In deep networks performing visual object recognition, a single layer of homogenous units exhibits a trade-off between various geometric transformations, while common network motifs involve beneficial geometrical changes to multiple geometric properties, suggesting the benefit of heterogeneity in neural populations [13]. More broadly, different brain regions relevant for distinct tasks may implement optimal neural geometry engendered by specific neuronal constraints. Given the vast heterogeneity of cell types, synaptic connectivity patterns, neuronal activation profiles, and sparsity levels, which biological properties constrain and shape the critical task-encoding geometry?

As geometric descriptions are general across task modalities, brain regions, and characteristic timescales, the neural population geometry approach may hold a key for unifying the descriptions of structure and function in biological and artificial neural networks across brain regions and computational levels.

# Conflict of interest statement

The authors declare no conflicts of interest, financial or otherwise.

# Acknowledgements

We thank Jacob Portes and David Clark for helpful comments. Research was supported by NSF NeuroNex Award (DBI-1707398), the Gatsby Charitable Foundation (GAT3708), and the Simons Collaboration for the Global Brain.

# Highlighted papers

- of special interest
- ●● of outstanding interest

**Chung S, Lee DD, Sompolinsky H. Classification and Geometry of General Perceptual Manifolds. Phys Rev X. 2018;8: 031003.**

●● The authors of this study generalize the theory of perceptron capacity of discrete points to categorical 'manifolds', for an arbitrary geometry of neural population responses for categories. 'Manifold capacity' is defined as a critical number of linearly separable category manifolds per neuron in a given neural representation. This manifold capacity can be formally expressed as a function of geometrical properties of the category manifolds, such as manifold dimension and manifold radius.

**Sohn H, Narain D, Meirhaeghe N, Jazayeri M. Bayesian Computation through Cortical Latent Dynamics. Neuron. 2019;103: 934–947.e5.**

● The authors introduce a geometrical approach for the population-level neural mechanisms underlying Bayesian behavior. In their study, prior-reflecting behavioral performance is affected by neural manifold geometry, such as curvature of dynamic trajectories, which has been warped by the statistics of the context-dependent prior.

**Russo AA, Bittner SR, Perkins SM, Seely JS, London BM, Lara AH, et al. Motor Cortex Embeds Muscle-like Commands in an Untangled Population Response. Neuron. 2018;97: 953–966.e8.**

●● The authors demonstrate population geometric measures for dynamic trajectories called 'tangling', which is found to be lower in primary motor cortex activity compared to EMG muscle activity. This suggests that motor cortex activity may be less driven by input dynamics, rendering better robustness to noise.

**Chaudhuri R, Gerçek B, Pandey B, Peyrache A, Fiete I: The intrinsic attractor manifold and population dynamics of a canonical cognitive circuit across waking and sleep.** *Nat Neurosci* **2019, 22:1512–1520.**

● This paper introduces a new method called Spline Parameterization for Unsupervised Decoding (SPUD), where the intrinsic dimensionality and the topology underlying high dimensional neural data are identified using a mathematical technique called persistent homology. Using this method, the authors show that the head direction of mice can be decoded from a one-dimensional ring structure in the neural population activity extracted from the post-subiculum and anterodorsal thalamus. This one-dimensional ring structure was also found during sleep, despite the lack of sensory input.

**Low RJ, Lewallen S, Aronov D, Nevers R, Tank DW. Probing variability in a cognitive map using manifold inference from neural dynamics. doi:10.1101/418939**

●● The authors introduce a topologically motivated method called Manifold Inference from Neural Dynamics (MIND) to characterize the neural activity during foraging and sound manipulation tasks, both of which involve a transition between behavioral states. For MIND, the distance between the nearby states was defined by transition probabilities, identifying dimensions relevant for topological maps underlying the task.

**Bernardi S, Benna MK, Rigotti M, Munuera J, Fusi S, Salzman CD. The Geometry of Abstraction in the Hippocampus and Prefrontal Cortex. Cell. 2020;183: 954–967.e21.**

●● The authors introduce a geometric measure called parallelism score, which characterizes the degree to which coding directions are parallel for different sets of training conditions and show that this geometry is related to cross-condition generalization performance, which characterizes how linear readouts can generalize across conditions. Neural representations conforming to this geometry were observed in the dorsolateral prefrontal cortex, anterior cingulate cortex, and the hippocampus in monkeys performing a serial reversal-learning task, as well as in neural networks.

# References

1.  Saxena S, Cunningham JP: **Towards the neural population doctrine**. *Curr Opin Neurobiol* 2019, **55**:103–111.

2.  Yuste R: **From the neuron doctrine to neural networks**. *Nat Rev Neurosci* 2015, **16**:487–497.

3.  Eichenbaum H: **Barlow versus Hebb: When is it time to abandon the notion of feature detectors and adopt the cell assembly as the unit of cognition?** *Neurosci Lett* 2018, **680**:88–93.

4.  Rumelhart DE, McClelland JL, Group PR, Others: *Parallel distributed processing*. IEEE Massachusetts; 1988.

5.  Seung HS, Lee DD: **Cognition. The manifold ways of perception**. *Science* 2000, **290**:2268–2269.

6.  DiCarlo JJ, Cox DD: **Untangling invariant object recognition**. *Trends Cogn Sci* 2007, **11**:333–341.

7.  DiCarlo JJ, Zoccolan D, Rust NC: **How does the brain solve visual object recognition?** *Neuron* 2012, **73**:415–434.

8.  Sohn H, Narain D, Meirhaeghe N, Jazayeri M: **Bayesian Computation through Cortical Latent Dynamics**. *Neuron* 2019, **103**:934–947.e5.

9.  Russo AA, Bittner SR, Perkins SM, Seely JS, London BM, Lara AH, Miri A, Marshall NJ, Kohn A, Jessell TM, et al.: **Motor Cortex Embeds Muscle-like Commands in an Untangled Population Response**. *Neuron* 2018, **97**:953–966.e8.

10. Low RJ, Lewallen S, Aronov D, Nevers R, Tank DW: **Probing variability in a cognitive map using manifold inference from neural dynamics**. 2018, doi:10.1101/418939.

11. Sadtler PT, Quick KM, Golub MD, Chase SM, Ryu SI, Tyler-Kabara EC, Yu BM, Batista AP: **Neural constraints on learning**. *Nature* 2014, **512**:423–426.

12. Chung S, Lee DD, Sompolinsky H: **Classification and Geometry of General Perceptual Manifolds**. *Phys Rev X* 2018, **8**:031003.

13. Cohen U, Chung S, Lee DD, Sompolinsky H: **Separability and geometry of object manifolds in deep neural networks**. *Nat Commun* 2020, **11**:746.

14. Jun JJ, Steinmetz NA, Siegle JH, Denman DJ, Bauza M, Barbarits B, Lee AK, Anastassiou CA, Andrei A, Aydın Ç, et al.: **Fully integrated silicon probes for high-density recording of neural activity**. *Nature* 2017, **551**:232–236.

15. Krizhevsky A, Sutskever I, Hinton GE: **Imagenet classification with deep convolutional neural networks**. *Adv Neural Inf Process Syst* 2012, **25**:1097–1105.

16. Le QV: **Building high-level features using large scale unsupervised learning**. In *2013 IEEE International Conference on Acoustics, Speech and Signal Processing*. . 2013:8595–8598.

17. Ramón y Cajal S: *Histology of the nervous system of man and vertebrates*. Oxford University Press, USA; 1995.

18. Sherrington CS: **Observations on the scratch-reflex in the spinal dog**. *The Journal of Physiology* 1906, **34**:1–50.

19. Seung HS, Sompolinsky H: **Simple models for reading neuronal population codes**. *Proc Natl Acad Sci U S A* 1993, **90**:10749–10753.

20. Brunel N, Nadal JP: **Mutual information, Fisher information, and population coding**. *Neural Comput* 1998, **10**:1731–1757.

21. Rigotti M, Barak O, Warden MR, Wang X-J, Daw ND, Miller EK, Fusi S: **The importance of mixed selectivity in complex cognitive tasks**. *Nature* 2013, **497**:585–590.

22. Yamins DLK, Hong H, Cadieu CF, Solomon EA, Seibert D, DiCarlo JJ: **Performance-optimized hierarchical models predict neural responses in higher visual cortex**. *Proc Natl Acad Sci U S A* 2014, **111**:8619–8624.

23. Khaligh-Razavi S-M, Kriegeskorte N: **Deep supervised, but not unsupervised, models may explain IT cortical representation**. *PLoS Comput Biol* 2014, **10**:e1003915.

24. Saxe A, Nelli S, Summerfield C: **If deep learning is the answer, what is the question?** *Nat Rev Neurosci* 2021, **22**:55–67.

25. Barrett DG, Morcos AS, Macke JH: **Analyzing biological and artificial neural networks: challenges with opportunities for synergy?** *Curr Opin Neurobiol* 2019, **55**:55–64.

26. Gallego JA, Perich MG, Miller LE, Solla SA: **Neural Manifolds for the Control of Movement**. *Neuron* 2017, **94**:978–984.

27. Jazayeri M, Afraz A: **Navigating the Neural Space in Search of the Neural Code**. *Neuron* 2017, **93**:1003–1014.

28. Cover TM: **Geometrical and Statistical Properties of Systems of Linear Inequalities with Applications in Pattern Recognition**. *IEEE Transactions on Electronic Computers* 1965, **EC-14**:326–334.

29. Gardner E: **The space of interactions in neural network models**. *J Phys A Math Gen* 1988, **21**:257.

30. Hénaff OJ, Goris RLT, Simoncelli EP: **Perceptual straightening of natural videos**. *Nat Neurosci* 2019, **22**:984–991.

31. Bernardi S, Benna MK, Rigotti M, Munuera J, Fusi S, Salzman CD: **The Geometry of Abstraction in the Hippocampus and Prefrontal Cortex**. *Cell* 2020, **183**:954–967.e21.

32. Stephenson C, Feather J, Padhy S, Elibol O, Tang H, McDermott J, Chung S: **Untangling in Invariant Speech Recognition**. In *Advances in Neural Information Processing Systems*. Edited by Wallach H, Larochelle H, Beygelzimer A, d\textquotesingle Alché-Buc F, Fox E, Garnett R. Curran Associates, Inc.; 2019:14391–14401.

33. Chung S, Lee DD, Sompolinsky H: **Linear readout of object manifolds**. *Phys Rev E* 2016, **93**:060301.

34. Mamou J, Le H, Rio MD, Stephenson C, Tang H, Kim Y, Chung S: **Emergence of Separable Manifolds in Deep Language Representations**. In *Proceedings of the 37th International Conference on Machine Learning, ICML 2020, 13-18 July 2020, Virtual Event.* . PMLR; 2020:6713–6723.

35. Kriegeskorte N: **Deep Neural Networks: A New Framework for Modeling Biological Vision and Brain Information Processing**. *Annu Rev Vis Sci* 2015, **1**:417–446.

36. Kell AJE, Yamins DLK, Shook EN, Norman-Haignere SV, McDermott JH: **A Task-Optimized Neural Network Replicates Human Auditory Behavior, Predicts Brain Responses, and Reveals a Cortical Processing Hierarchy**. *Neuron* 2018, **98**:630–644.e16.

37. Schrimpf M, Blank I, Tuckute G, Kauf C, Hosseini EA, Kanwisher N, Tenenbaum J, Fedorenko E: **Artificial Neural Networks Accurately Predict Language Processing in the Brain**. *Cold Spring Harbor Laboratory* 2020, doi:10.1101/2020.06.26.174482.

38. Froudarakis E, Cohen U, Diamantaki M, Walker EY, Reimer J, Berens P, Sompolinsky H, Tolias AS: **Object manifold geometry across the mouse cortical visual hierarchy**. *Cold Spring Harbor Laboratory* 2020, doi:10.1101/2020.08.20.258798.

39. Chung S, Dapello J, Cohen U, DiCarlo J, Sompolinsky H: **Separable Manifold Geometry in Macaque Ventral Stream and DCNNs**. In *Computational and Systems Neuroscience.* . 2020.

40. Balasubramanian M: **The Isomap Algorithm and Topological Stability**. *Science* 2002, **295**:7a–7.

41. Roweis ST: **Nonlinear Dimensionality Reduction by Locally Linear Embedding**. *Science* 2000, **290**:2323–2326.

42. van der Maaten L: **Visualizing Data using t-SNE**. 2008,

43. Borg I, Groenen PJF: *Modern Multidimensional Scaling: Theory and Applications*. Springer Science & Business Media; 2005.

44. Gigante S, Charles AS, Krishnaswamy S, Mishne G: **Visualizing the PHATE of Neural Networks**. In *Advances in Neural Information Processing Systems*. . 2019.

45. McInnes L, Healy J, Melville J: **UMAP: Uniform Manifold Approximation and Projection for Dimension Reduction**. *arXiv [statML]* 2018,

46. Chaudhuri R, Gerçek B, Pandey B, Peyrache A, Fiete I: **The intrinsic attractor manifold and population dynamics of a canonical cognitive circuit across waking and sleep**. *Nat Neurosci* 2019, **22**:1512–1520.

47. Giusti C, Pastalkova E, Curto C, Itskov V: **Clique topology reveals intrinsic geometric structure in neural correlations**. *Proceedings of the National Academy of Sciences* 2015, **112**:13455–13460.

48. Dabaghian Y, Brandt VL, Frank LM: **Reconceiving the hippocampal map as a topological template**. *Elife* 2014, **3**:e03476.

49. Low RJ, Lewallen S, Aronov D, Nevers R, Tank DW: **Probing variability in a cognitive map using manifold inference from neural dynamics**. [date unknown], doi:10.1101/418939.

50. Nieh EH, Schottdorf M, Freeman NW, Low RJ, Lewallen S, Koay SA, Pinto L, Gauthier JL, Brody CD, Tank DW: **Geometry of abstract learned knowledge in the hippocampus**. *Nature* 2021, **595**:80–84.

51. Russo AA, Khajeh R, Bittner SR, Perkins SM, Cunningham JP, Abbott LF, Churchland MM: **Neural Trajectories in the Supplementary Motor Area and Motor Cortex Exhibit Distinct Geometries, Compatible with Different Classes of Computation**. *Neuron* 2020, **107**:745–758.e6.

52. Hopfield JJ: **Neural networks and physical systems with emergent collective computational abilities**. *Proc Natl Acad Sci U S A* 1982, **79**:2554–2558.

53. Amari S-I: **Neural theory of association and concept-formation**. *Biological Cybernetics* 1977, **26**:175–185.

54. Seung HS: **How the brain keeps the eyes still**. *Proc Natl Acad Sci U S A* 1996, **93**:13339–13344.

55. Kopell N, Ermentrout GB: **Symmetry and phaselocking in chains of weakly coupled oscillators**. *Commun Pure Appl Math* 1986, **39**:623–660.

56. Shenoy KV, Sahani M, Churchland MM: **Cortical control of arm movements: a dynamical systems perspective**. *Annu Rev Neurosci* 2013, **36**:337–359.

57. Mante V, Sussillo D, Shenoy KV, Newsome WT: **Context-dependent computation by recurrent dynamics in prefrontal cortex**. *Nature* 2013, **503**:78–84.

58. Jazayeri M, Ostojic S: **Interpreting neural computations by examining intrinsic and embedding dimensionality of neural activity**. *Curr Opin Neurobiol* 2021, **70**:113–120.

59. Stringer C, Pachitariu M, Steinmetz N, Carandini M, Harris KD: **High-dimensional geometry of population responses in visual cortex**. *Nature* 2019, **571**:361–365.

60. Ehrlich DB, Murray JD: **Geometry of neural computation unifies working memory and planning**. *bioRxiv* 2021, doi:10.1101/2021.02.01.429156.

61. Gallego JA, Perich MG, Naufel SN, Ethier C, Solla SA, Miller LE: **Cortical population activity within a preserved neural manifold underlies multiple motor behaviors**. *Nat Commun* 2018, **9**:4233.

62. Kobak D, Pardo-Vazquez JL, Valente M, Machens C, Renart A: **State-dependent geometry of population activity in rat auditory cortex**. *Elife* 2019, doi:10.1101/501338.

63. Okazawa G, Hatch CE, Mancoo A, Machens CK, Kiani R: **Representational geometry of perceptual decisions in the monkey parietal cortex**. *Cell* 2021, **184**:3748–3761.e18.